\documentclass[aps,preprint]{revtex4}
\usepackage{amsfonts}
\usepackage{amsmath}
\usepackage{amssymb}
\usepackage{epsf}
\usepackage{epsfig}
\usepackage{graphicx}
\usepackage{rotating}
\usepackage{boxedminipage}

\input{tcilatex}

\begin{document}

\title{MODELLING INVESTMENT IN ARTIFICIAL STOCK MARKETS: ANALYTICAL AND
NUMERICAL RESULTS}
\author{Roberto da Silva}
\affiliation{Departamento de Inform\'{a}tica Te\'{o}rica, Instituto de Inform\'{a}tica,
Universidade Federal do Rio Grande do Sul. \\
Av. Bento Gon\c{c}alves, 9500, CEP 90570-051, Porto Alegre, RS, Brazil}
\email{rdasilva@inf.ufrgs.br}
\author{Alexandre Tavares Baraviera}
\affiliation{Instituto de Matem\'{a}tica, Universidade Federal do Rio Grande do Sul, Av.
Bento Gon\c{c}alves, 9500, CEP 91509-900, Porto Alegre, RS, Brazil}
\email{baravi@mat.ufrgs.br}
\author{Silvio Renato Dahmen}
\affiliation{Instituto de F\'{\i}sica, Universidade Federal do Rio Grande do Sul, Av.
Bento Gon\c{c}alves, 9500, CEP 91501-970, Porto Alegre, RS, Brazil}
\email{dahmen@if.ufrgs.br}

\begin{abstract}
In this article we study the behavior of a group of economic agents in the
context of cooperative game theory, interacting according to rules based on
the Potts Model with suitable modifications. Each agent can be thought of as
belonging to a chain, where agents can only interact with their nearest
neighbors (periodic boundary conditions are imposed). Each agent can invest
an amount $\sigma _{i}=0,...,q-1$. Using the transfer matrix method we study
analytically, among other things, the behavior of the investment as a
function of a control parameter (denoted $\beta $) for the cases $q=2$ and $3
$. For $q>3$ numerical evaluation of eigenvalues and high precision
numerical derivatives are used in order to assess this information.

\textbf{keywords:} Game Theory; statistical mechanics; transfer matrix
methods.
\end{abstract}

\maketitle

\setlength{\baselineskip}{0.7cm}

\section{Introduction}

Consider a game where N players can invest their money (up to some upper
limit) on some public--fund asset. The fund manager, as a rule, doubles the
amount of money received and divides it equally among all N investors.
Depending on the total amount each player invested, some might end up making
a profit while others may lose money. Assume that players have no
information whatsoever about their co-players' moves. The question is: what
is the best move a player can make? If we adopt one of the tenets of
classical game theory, namely that players are completely rational, then
there are two possible solutions to the problem which maximize profit (when
all invest the maximum amount possible, thus doubling their initial capital)
or minimize losses (no one invests anything). In the real world however
people are not rational in the sense of classical game theory and factors as
expectations about the behavior of other players or some sort of insider
information might play a role when deciding how to invest.

The irrationality of market agents is one of a myriad of factors which
account for the high complexity of financial markets and the difficulty in
modelling them. Markets may also be affected by political turmoil,
unseasonable weather variations and the like. In the past few years models
have been introduced in order to throw some light into the behavior of
markets, mainly with aims at forecasting long-term behavior see for example 
\cite{Br1993,Bl1993}. These models, which have the advantage of being either
analytically or numerically treatable and usually use some kind of data
input from real markets, are nonetheless unable to take into account the
human factor in decision-making scenarios. To circumvent these difficulties
a new approach, inspired on the ideias of statistical mechanics has been
suggested \cite{D1999}, where one extends the set of causal factors in
decision--making scenarios from the individual--specific to group
determinants of behavior: players' decisions are not market--mediated but
rely on group--level influences.

With these ideias in mind our aim in this work is to extend the model for
the game discussed above through the introduction of cooperation, \textit{%
i.e.} we allow agents to have partial information on the decision of its
immediate neighbors, in a way to be described precisely in what follows.
Furthermore, we allow for some kind of randomness, the only thing known a
priori being the probability of some decision, and not the decision itself.
In this way we hope to describe the average behavior of a large group of
agents without entering in the details of a realistic (and certainly very
difficult) theory on psychological state of each agent.

Our main interest will be to see how is the average behavior of a (large)
group of cooperative economic agents. Each agent, labelled by the index $i$,
is allowed to invest an amount $\sigma_{i}$. Before the investment, the
agent $i$ exchange information with agent $i+1$ (defined as the neighbor of $%
i$; this concept is symmetric, i.e., $i$ is also neighbor of $i+1$). Based
on that information agents make decisions as to how much they will invest
according to some probability distribution parameterized in terms of a two
real variables: $J$ and $\beta$. The first measures how strongly people
interact with each other (group--level influence) while the later is a
measure of how strongly a player might deviate from the group. To model this
we choose a suitably defined function which measures the probability of $i$
investing $\sigma _{i}$ given that $i+1$ would like to invest $\sigma _{i+1}$%
. In a way to be precisely formalized later, $\beta$ allows us to change the
expected behavior of each agent.

The paper is organized as follows: In section $2$ we explain the model and
make the connection with statistical mechanics. Using the standard transfer
matrix technique\cite{H1987} we analyze in section $3$ some integrable cases
($q=2$ and $q=3$) in order to gain information about how the average
investment changes as a function of $\beta $. In section $4$ we numerically
evaluate the evolution for $q\geq 4$. We introduce a method for calculating
the investment using derivatives of the biggest eigenvalue, which is based
on the use of 5 points in a graphic. We finish the paper with some
conclusions and perspectives.

\section{Formulating the problem: the Potts Model}

We consider an ensemble of $N$ agents, where each can invest an amount $%
\sigma_i \in \{0, 1, \dots, q-1\}$, $q$ a fixed integer. This restriction on 
$q$ has been made for the sake of clarity. The methods employed can be
easily generalized to the case where $\sigma_i \in \{ d_0, \ldots, d_{q-1} \}
$, the $d_i$'s being arbitrary real numbers.

The families of conditional probabilities, \textit{i.e.} the probability
that $i+1$ would invest $\sigma_{i+1}$ given that $i$ invested $\sigma_i$
are chosen as 
\begin{equation}
P(\sigma _{i+1}|\sigma _{i})=c\exp \left[ -\beta \cdot J(\sigma _{i})\cdot
\delta _{\sigma _{i},\sigma _{i+1}}\right] \;.  \label{conditional}
\end{equation}
Our motivation for this particular choice comes from physics, where this
probability is interpreted as that of two variables (called classical spins) 
$\sigma_i$ and $\sigma_{i+1}$ being equal or having different values. It
depends on two physical parameters: $J(\sigma_i)$ is the so--called
interaction strength. This is in most cases a material--dependent quantity
and accounts for the different collective properties materials may exhibit
(ferromagnetic, antiferromagnetic, etc.). $\beta$ is proportional to the
inverse temperature $T^{-1}$ and brings about entropic effects (ordered
states for low temperatures and disorder for high temperatures). Being
proportional to a temperature, in physical systems $\beta$ is always non
negative, and we likewise assume our $\beta\geq 0$.

In general $J$ and $\beta$ can be seen as competing terms: $J$ is associated
to the energy cost of a given spin configuration. For $J>0$ ($<0$) a
configuration where spins are equal (different) has a higher energy than the
opposite configuration, which means that it is energetically more favorable
to be non--magnetic (or ferromagnetic); on the other hand, the temperature $%
\beta^{-1}$ tends to destroy magnetic order. Transposing these ideas to the
financial context can say that $J(\sigma_i)$ measures how strongly people
respond to their neighbors' moves, that is if they are susceptible to the
influence of other players or not. In this sense it models distinct profiles
of investors, which can go from agressive (does not go along ``with the
pack") to conservative (does what others do). $\beta$ is a measure of the
strength of individual response and independent of what others do. Social
scientists refer to this term as the ``individual--specific random" and
``unobservable" (from the point of view of the modeler) since it is
associated to personal beliefs \cite{D1999}.

With the neighbor-to-neighbor interaction rule introduced above we can
describe the behavior of the whole group: The first important quantity which
needs to be defined is the joint probability density (j.p.d), $P(\sigma
_{1},\sigma_{2},...,\sigma _{N}) $ of a particular investment configuration $%
\sigma \equiv(\sigma_{1},\sigma_{2},...,\sigma _{N})$ of a group. The
quantity invested is defined through 
\begin{equation}
L(\sigma )=\sigma _{1}+\sigma _{2}+...+\sigma _{N}
\label{investimento_total}
\end{equation}
and we would like to obtain the average value of $L$.

To calculate the j.p.d we may adopt a recursive formulation without any loss
of generality: we take the investment of the first agent to be exactly $%
\sigma _{1}$, i.e., $P(\sigma _{1})=1$ and from that derive the quantity we
want. With this ``boundary condition" on $P(\sigma _{1})$ we have the
following theorem:

\begin{theorem}
The probability distribution $P(\sigma _{1},\sigma _{2},...,\sigma
_{N})$ can be written as a product form
\begin{equation}
P(\sigma _{1},\sigma _{2},...,\sigma
_{N})=\prod\limits_{i=1}^{N-1}P(\sigma _{i+1}|\sigma _{i})
\label{product}
\end{equation}%
with $P(\sigma _{i}|\sigma _{i-1},...,\sigma
_{1})=P(\sigma _{i}|\sigma _{i-1})$ for $i=1,...,N$.
\end{theorem}

\begin{proof}
>From the definition
\begin{equation}
P(\sigma _{N}|\sigma _{1},\sigma _{2},...,\sigma
_{N-1})=\frac{P(\sigma _{1},\sigma _{2},...,\sigma _{N})}{P(\sigma
_{1},\sigma _{2},...,\sigma _{N-1})}  \label{conditionalII}
\end{equation}%
Considering the hypothesis $P(\sigma _{N}|\sigma _{1},\sigma
_{2},...,\sigma _{N-1})=P(\sigma _{N}|\sigma _{N-1})$ we thus have
\begin{equation}
P(\sigma _{1},\sigma _{2},...,\sigma _{N})=P(\sigma _{1},\sigma
_{2},...,\sigma _{N-1})P(\sigma _{N+1}|\sigma _{N})\; .
\label{conditionalIII}
\end{equation}%
and applying this recursively:%
\begin{equation*}
\begin{array}{lll}
P(\sigma _{1},\sigma _{2},...,\sigma _{N}) & = & P(\sigma
_{1},\sigma
_{2},...,\sigma _{N-1})P(\sigma _{N}|\sigma _{N-1}) \\
&  &  \\
& = & P(\sigma _{1},\sigma _{2},...,\sigma _{N-2})P(\sigma
_{N-1}|\sigma
_{N-2})P(\sigma _{N}|\sigma _{N-1}) \\
&  &  \\
& = & P(\sigma _{1})\prod\limits_{i=1}^{N-1}P(\sigma _{i+1}|\sigma
_{i})=\prod\limits_{i=1}^{N-1}P(\sigma _{i+1}|\sigma _{i})%
\end{array}%
\end{equation*}
\end{proof}
According to equations (\ref{conditional}) and (\ref{product}) we have 
\begin{equation*}
P(\sigma _{1},\sigma _{2},...,\sigma _{N})=c^{N}\exp \left[ -\beta
\sum_{i=1}^{N-1}J(\sigma _{i})\cdot \delta _{\sigma _{i},\sigma _{i+1}}%
\right]
\end{equation*}%
where $c^{N}$ is the normalization constant defined before and such that 
\begin{equation*}
\sum\limits_{\sigma _{1}=d_{0}}^{d_{q-1}}\sum\limits_{\sigma
_{2}=d_{0}}^{d_{q-1}}...\sum\limits_{\sigma _{N}=d_{0}}^{d_{q-1}}P(\sigma
_{1},\sigma _{2},...,\sigma _{N})=1\qquad .
\end{equation*}

\subsection{Investment formulas and the Potts model}

The Potts hamiltonian of $N$ interacting spins under the action of a
magnetic field $D$ is given by \cite{W1973} 
\begin{equation}  \label{pottshamiltonian}
H\left( \sigma \right) = \sum\limits_{\left\langle i,j\right\rangle}J(\sigma
_{i}) \delta_{\sigma _{i},\sigma _{j}} +D\sum\limits_{i=1}^{N} \sigma_{i}.
\end{equation}
The fact that the total investment $L$ (\ref{investimento_total}) is
mathematically the same as the magnetization of the Potts hamiltonian (\ref%
{pottshamiltonian}) means that we may directly transpose the techniques and
ideas of statistical mechanics into the financial scenario.

The probability density function for the system to present a specific
investment value $L$ is given by 
\begin{equation}
P(L)=\left\{ 
\begin{array}{lll}
Z_{N}^{-1}\exp \left[ -\beta H(\sigma )\right] & \;\;\;\;\; {\mbox{if}} & 
\;\; L=\sum\nolimits_{i=1}^{N}\sigma _{i} \\ 
&  &  \\ 
0 & {\mbox{otherwise}} & 
\end{array}%
\right.  \label{prob_distrib}
\end{equation}%
where $Z_{N}(\beta )$, the normalization constant, is a sum over all
possible configurations 
\begin{equation}
Z_{N}(\beta )=\sum\limits_{\sigma _{j}=0,...,q-1}\exp \left[ -\beta H\left(
\sigma \right) \right] .  \label{particion_function_I}
\end{equation}
and is known as the partition function.

Our aim is to describe how the investment depends on $\beta $, given a fixed
set of parameters $J(1),J(2),...,J(N)$. For this purpose, let us consider
the expected value of $L$ according to the distribution (\ref{prob_distrib}%
), \textit{i.e.} 
\begin{equation*}
\left\langle L\left( \sigma \right) \right\rangle =\frac{1}{Z}%
\sum\limits_{\sigma _{j}=0,...,q-1}\left( \sum\limits_{i=1}^{N}\sigma
_{i}\right) \exp \left[ -\beta H\left( \sigma \right) \right] .
\end{equation*}

For the sake of those not familiar with the methods of statistical
mechanics, we briefly discuss how in our \textit{analogy between spin
systems and economic games} quantities of interest can be calculated:

\begin{enumerate}
\item The term $D\sum\limits_{i=1}^{N}\sigma _{i}$ is introduced for
convenience since \textbf{investment per capita} is calculated through the
formulae 
\begin{equation*}
l(\beta )=\frac{1}{N}\left\langle L(\beta )\right\rangle =\left. -\frac{1}{%
\beta N}\frac{\partial }{\partial D}\log Z_{N}(\beta )\right\vert _{D=0},
\end{equation*}%
which is the analogue in statistical mechanics to the average spontaneous
magnetization. It clearly obeys the inequality $0\leq l\leq q-1$.

\item As previously discussed, investment depends essentially on the way how
neighbors cooperate, \textit{i.e.} the distribution $J(0),J(1),...$.$J(q-1)$
defines the kind of profiles investors have.
\end{enumerate}

The first question one might ask would be: what is the behavior of the per
capita investment $l(\beta )$ at $\beta=0$ and $\beta \rightarrow \infty$ ?
One may describe these limits in a straightforward manner: 
\begin{theorem}
The per capita investment is such that

$l(0)=\frac{q-1}{2};$

$l(\infty )=\sigma _{min}$ if $J_{min}$ satisfies the inequality
$J_{min} < J_{k}$ for all $k=0, \ldots, min-1, min+1, \ldots,
q-1$.
\end{theorem}
\begin{proof}
The probability of a given state $\sigma =\left\{ \sigma
_{i}\right\} _{i=1}^{N}$ is
\begin{equation*}
P(\sigma )=\frac{e^{-\beta H(\sigma )}}{Z_{N}(\beta )};
\end{equation*}%
 For $\beta =0$ all states are
equiprobable since, from the equation above, the probability does
not depend on $\sigma$; hence, the probability of each state is
$1/q^N$. For simplicity we assume that $q^N$ is even, the odd case
being left to the reader.
The per capita investment can be written as
\[
   l(0)= \frac{1}{N q^N} \sum_{i=0}^{N(q-1)} (\mbox{number of states with sum $=i$}) i
\]
We can see that the number of states corresponding to the sum $i$ is the same as
the number corresponding to the sum $N(q-1)-i$. Then,
\[
   l(0)= \frac{1}{N q^N} \sum_{i=0}^{N(q-1)/2}
(\mbox{number of states with sum $=i$}) N(q-1) =
\]
\[
\frac{1}{N q^N} N(q-1) \frac{q^N}{2} = \frac{q-1}{2}
\]
since the number of states with sum between $0$ and $N(q-1)/2$
correspond to $(q^N)/2$.

 For arbitrary values of $\beta $ let us call $\sigma
_{M}=\left\{ \sigma _{min} \right\} _{i=1}^{N}$. Then, the
probability of any state is
\begin{equation*}
P(\sigma )=\frac{e^{-\beta H(\sigma )}}{e^{-\beta H(\sigma
_{M})}\left( 1+\sum\nolimits_{\sigma ^{\prime }\neq \sigma
_{M}}\frac{e^{-\beta H(\sigma ^{\prime })}}{e^{-\beta H(\sigma
_{M})}}\right) }
\end{equation*}%
In the limit $\beta \rightarrow \infty $, $P(\sigma )=0$
except for the state $\sigma _{M}$, where $P(\sigma _{M})=1.$
\end{proof}
Notice that the case where $\min_i{J_i}$ is reached in more than one point
is not covered by the theorem.

In the next section we give a case by case description of investment as a
function of $\beta $. To do this we adapt transfer matrix method to our
model. Contrary to spin systems, in the present problem nontrivial behavior
appears, and this might lead to interesting new possibilities in the
scenario of economic games. 

\section{Analytical cases}

\subsection{The two state model: $q=2$ (Ising Model)}

We start by considering the partition function (\ref{particion_function_I})

\begin{equation}
\begin{array}{llll}
Z & = & \sum\limits_{\sigma _{1}=0,1}\cdot \cdot \cdot \sum\limits_{\sigma
_{N}=0,1}\exp \left( -\beta \sum\limits_{i=1}^{N}J(\sigma _{i})\delta
_{\sigma _{i},\sigma _{i+1}}-\frac{\beta }{2}D\sum\limits_{i=1}^{N}(\sigma
_{i}+\sigma _{i+1})\right) &  \\ 
&  &  &  \\ 
& = & \sum\limits_{\sigma _{1}=0,1}\cdot \cdot \cdot \sum\limits_{\sigma
_{N}=0,1}\prod\limits_{i=1}^{N}\exp \left( -\beta J(\sigma _{i})\delta
_{\sigma _{i},\sigma _{i+1}}-\frac{\beta }{2}D(\sigma _{i}+\sigma
_{i+1})\right) &  \\ 
&  &  &  \\ 
& = & Tr\;M^{N} & 
\end{array}
\label{partition_function}
\end{equation}%
where $\sigma _{N+1}=\sigma _{1}$ (periodic boundary conditions are assumed)
and the transfer matrix $M$ is given by 
\begin{equation*}
M=\left( 
\begin{array}{cc}
\exp \left( -\beta J_{0}\right) & \exp \left( -(\beta /2)D\right) \\ 
\exp \left( -(\beta /2)D\right) & \exp \left( -\beta J_{1}-\beta D\right)%
\end{array}%
\right)
\end{equation*}%
It is possible to diagonalize $M$ and write (\ref{partition_function}) in
terms of eigenvalues $\lambda _{1}$ and $\lambda _{2}$ of $M$ in a simple
way 
\begin{equation}
Z_{N}(\beta ,D)=\lambda _{1}^{N}+\lambda _{2}^{N}
\end{equation}%
These eigenvalues are given by 
\begin{equation*}
\lambda _{1,2}(\beta)=\frac{1}{2}\left[ \left( e^{-\beta J_{0}}+e^{-D\beta
-\beta J_{1}}\right) \pm \sqrt{\Theta }\right]
\end{equation*}%
with 
\begin{eqnarray*}
\Theta &=&4e^{-D\beta }+e^{-2(\beta J_{0})}+e^{-2\beta (D+J_{1})}-2e^{-\beta
(D+J_{0}+J_{1})} \\
J_{i} &=&J(i)\qquad \qquad \qquad \qquad i=0,1\;.
\end{eqnarray*}%
One can now write 
\begin{equation*}
\begin{array}{lllll}
l(\beta ) & = & -\frac{1}{\beta N}\left. \frac{\partial }{\partial D}\log
\left( {\lambda }_{1}^{N}{+\lambda }_{2}^{N}\right) \right\vert _{D=0} &  & 
\\ 
&  &  &  &  \\ 
& = & -\left. \frac{1}{\beta N}\frac{1}{(\lambda _{1}^{N}+\lambda _{2}^{N})}%
\left( N\lambda _{1}^{N-1}\frac{\partial }{\partial D}\lambda _{1}+N\lambda
_{2}^{N-1}\frac{\partial }{\partial D}\lambda _{2}\right) \right\vert _{D=0}
&  &  \\ 
&  &  &  &  \\ 
& = & -\left. \frac{1}{\beta }\frac{1}{\lambda _{1}}\frac{1}{(1+(\lambda
_{2}/\lambda _{1})^{N})}[\frac{\partial }{\partial D}\lambda _{1}+(\frac{%
\lambda _{2}}{\lambda _{1}})^{N-1}\frac{\partial }{\partial D}\lambda
_{2}]\right\vert _{D=0} &  & 
\end{array}%
\end{equation*}%
In the limit of a large group of agents  ($N\rightarrow \infty )$ the
expression above converges to 
\begin{equation}
l(\beta )=-\left. \frac{1}{\beta }\frac{1}{\lambda _{1}}\frac{\partial }{%
\partial D}\lambda _{1}\right\vert _{D=0}  \label{derlambda1}
\end{equation}%
An explicit calculation gives 
\begin{equation*}
l(\beta ,J_{0},J_{1})=\frac{-e^{\beta (J_{1}-J_{0})}+2e^{2\beta J_{1}}+1+%
\sqrt{\Delta }}{4e^{2\beta J_{1}}-2e^{\beta (J_{1}-J_{0})}+e^{2\beta
(J_{1}-J_{0})}+e^{\beta (J_{1}-J_{0})}\sqrt{\Delta }+\sqrt{\Delta }+1}
\end{equation*}%
where 
\begin{equation}
\Delta (\beta ,J_{0},J_{1})=4e^{2\beta J_{1}}-2e^{\beta J_{1}}e^{-\beta
J_{0}}+e^{-2\beta J_{0}}e^{2\beta J_{1}}+1\;.
\end{equation}%
A few remarks can be drawn from these equations: 

\begin{enumerate}
\item With the explicit expressions of the eigenvalues one can easily show
that 
\begin{equation*}
l(\beta =0)=\frac{1}{2}
\end{equation*}
independently of the values of $J_{0}$ and $J_{1}$ (as expected from the
last theorem).

\item When  $J_{0}=J_{1}$ one has $\Delta =4e^{2\beta J_0}$ and
\end{enumerate}

\begin{equation*}
l(\beta )=1/2\;\; \mbox{for all} \;\; \beta \in \mathbb{R}
\end{equation*}%
The other important limit $l(\beta \rightarrow \infty )$ has an explicit $%
J_{i}$ dependence that can be summarized below: 
\begin{equation*}
l(\beta \rightarrow \infty )=\left\{ 
\begin{array}{ccccc}
\frac{1}{2} &  & {\mbox{if}}J_{0}>J_{1}>0 & {\mbox{or}} & J_{1}>J_{0}>0 \\ 
&  &  &  &  \\ 
1 &  & {\mbox{if}}J_{1}<J_{0}<0 & {\mbox{or}} & J_{1}<0\;\; {\mbox{and}}\;\;
J_{0}>0 \\ 
&  &  &  &  \\ 
0 &  & {\mbox{if}}J_{0}<J_{1}<0\  & {\mbox{or}} & J_{1}>0\;\; {\mbox{and}}%
\;\; J_{0}<0%
\end{array}%
\right.
\end{equation*}

This result follows from the expression for $l(\beta, J_0, J_1)$ above. At
this point we would like to make an important comment on the mean value of $%
J_{i}$. More specifically we are interested in the positivity or negativity
of $J_i$ for this tells us to what extent agents are cooperating and the
probability of their making investments.For example, if $J_{0}>0$ and $%
J_{1}<0$, the probability of two neighbors investing the same $\sigma _{i}=1$
is greater than the cojoint investment of $\sigma _{i}=0$.

\begin{figure}[th]
\centerline{\psfig{file=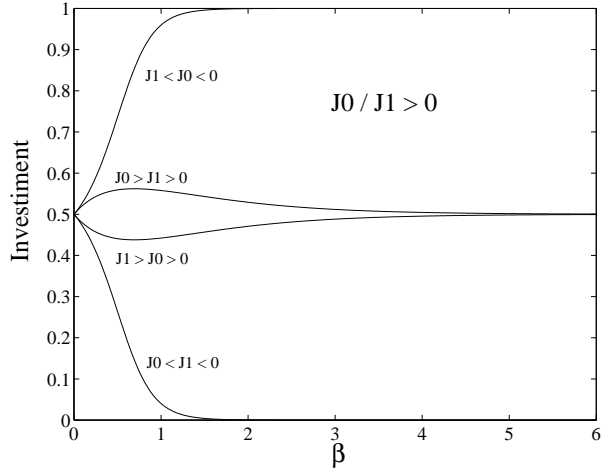,width=8cm}} \vspace*{8pt}
\caption{{}Investment as function of $\protect\beta$. Case $J_{0}/J_{1}>0$, $%
q=2$.}
\label{figure 1}
\end{figure}

\begin{figure}[th]
\centerline{\psfig{file=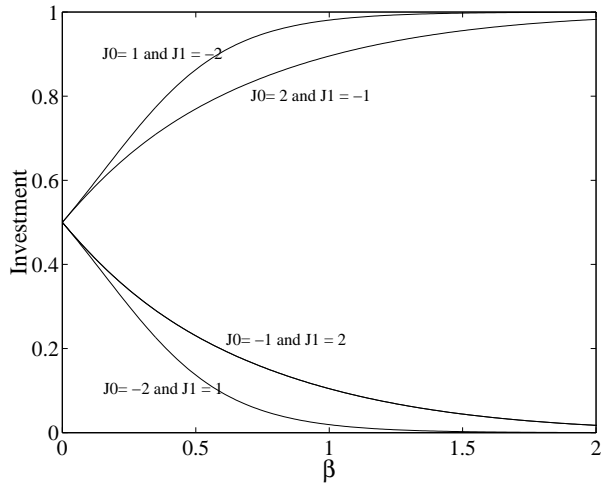,width=8cm}} \vspace*{8pt}
\caption{{}Investment as function of $\protect\beta$ to $q=2$, case $%
J_{0}/J_{1}<0$}
\label{figure 2}
\end{figure}

This behavior can be better seen in Figs. \ref{figure 1} and \ref{figure 2}.
The first one depicts investment as a function of $\beta$ for the ratio $%
J_{0}/J_{1}>0$. In the second figure the sign of this ratio is reversed.
These clearly show how investment behavior (how agents cooperate) is
drastically modified as a function of the profit and depends not only on 
\textbf{cash flow }(cash receipts minus cash payments over a given period of
time).

\subsection{The three state model: $q=3$}

In this case we have the following transfer matrix: 
\begin{equation*}
M=\left( 
\begin{array}{ccc}
e^{-\beta J_{0}} & e^{-\beta D/2} & e^{-\beta D} \\ 
e^{-\beta D/2} & e^{-\beta (J_{1}+D)} & e^{-3\beta D/2} \\ 
e^{-\beta D} & e^{-3\beta D/2} & e^{-\beta (J_{2}+2D)}%
\end{array}%
\right)
\end{equation*}

The problem of computing eigenvalues is analytically tractable for general
values of set $\{J_{0},J_{1},J_{2}\}$ but the expressions obtained are
generally very difficult and do not improve our understanding of the
problem. However some interesting sub-cases can be considered:

\subsubsection{Case 1: $J_{0}=J_{1}=0$ and $J_{2}=J$}

\label{subsection421} In this case by making the change of variables 
\begin{equation*}
x=e^{-\beta J} \;\;{\mbox{and}}\;\; y=e^{-\beta D/2}
\end{equation*}%
we arrive at 
\begin{equation}
M=\left( 
\begin{array}{ccc}
1 & y & y^{2} \\ 
y & y^{2} & y^{3} \\ 
y^{2} & y^{3} & xy^{4}%
\end{array}%
\right)  \label{j0=j1=0 e j2=j}
\end{equation}

One may clearly see that in this matrix the second row is obtained from the
first through multiplication by $y$. Hence det$M=0$ and therefore $M$ admits 
$0$ as an eigenvalue. One may thus compute analytically the other
eigenvalues by solving the equation 
\begin{equation*}
\lambda ^{2}-(1+y^{2}+xy^{4})\lambda +(x-1)(y^{4}+y^{6})=0,
\end{equation*}%
where the largest eigenvalue is 
\begin{equation*}
\lambda =\frac{1}{2}\left( 1+y^{2}+xy^{4}+\sqrt{\Delta }\right)
\end{equation*}%
Here we have 
\begin{equation*}
\Delta =2y^{2}+5y^{4}+4y^{6}-2xy^{4}-2xy^{6}+x^{2}y^{8}+1\Rightarrow \Delta
(D=0)=12-4x+x^{2}
\end{equation*}%
and so 
\begin{equation*}
\begin{array}{lllll}
l(\beta ) & = & -\left. \frac{1}{\beta \lambda }\frac{\partial \lambda }{%
\partial y}\frac{\partial y}{\partial D}\right\vert _{D=0} &  &  \\ 
&  &  &  &  \\ 
& = & \frac{1}{\left( 2+x+\sqrt{12-4x+x^{2}}\right) }\left[ 1+2x+\frac{%
12-5x+2x^{2}}{\sqrt{12-4x+x^{2}}}\right] &  &  \\ 
&  &  &  &  \\ 
& = & \frac{1}{\left( 2+e^{-\beta J}+\sqrt{12-4e^{-\beta J}+e^{-2\beta J}}%
\right) }\left[ 1+2e^{-\beta J}+\frac{12-5e^{-\beta J}+2e^{-2\beta J}}{\sqrt{%
12-4e^{-\beta J}+e^{-2\beta J}}}\right] &  & 
\end{array}%
\end{equation*}
One may observe that two cases follow from theorem 3.1: $\beta=0$ and $\beta
\to \infty$ when $J <0$. We have $l(0)=1$ and if $J < 0$, $l(\beta
\rightarrow \infty)=2$. But with the expression for $l$ we can also obtain
results beyond the range of the theorem, for example, when $J>0$. In this
case $l(\beta \rightarrow \infty )=\frac{1}{2+\sqrt{12}}\left[ 1+\sqrt{12}%
\right] $ $=\allowbreak 0.816\,97\cdots $\thinspace.

\subsubsection{Case 2: $J_{0}=J_{2}=0$ and $J_{1}=J$}

In this case we have%
\begin{equation*}
M=\left( 
\begin{array}{ccc}
1 & y & y^{2} \\ 
y & xy^{2} & y^{3} \\ 
y^{2} & y^{3} & y^{4}%
\end{array}%
\right)
\end{equation*}
The discussion is analogous to that of section (\ref{subsection421}). The
largest eingenvalue is given by 
\begin{equation*}
\lambda =\frac{1}{2}\left[ (y^{4}+xy^{2}+1)+\sqrt{\Delta }\right] ,
\end{equation*}%
with 
\begin{equation*}
\Delta =4y^{2}+2y^{4}+4y^{6}+y^{8}-2xy^{2}-2xy^{6}+x^{2}y^{4}+1\;.
\end{equation*}%
A straighforward calculation gives 
\begin{equation*}
l(\beta)=1
\end{equation*}
Thus the investment is independent of the parameter $J_{1}$ when $%
J_{0}=J_{2}=0.$

\subsubsection{Case 3: $J_{0}=J$ and $J_{1}=J_{2}=0$}

Now $M$ takes the form 
\begin{equation*}
M=\left( 
\begin{array}{ccc}
x & y & y^{2} \\ 
y & y^{2} & y^{3} \\ 
y^{2} & y^{3} & y^{4}%
\end{array}%
\right)
\end{equation*}
As before we have 
\begin{equation*}
\lambda =\frac{1}{2}\left[ x+y^{2}+y^{4}+\sqrt{\Delta }\right]
\end{equation*}%
with 
\begin{equation*}
\Delta
=x^{2}+4y^{2}+5y^{4}+2y^{6}+y^{8}-2xy^{2}-2xy^{4}=x^{2}+12-2x-2x=12+x^{2}-4x
\end{equation*}

For $l(\beta)$ we have the following expression 
\begin{equation*}
l(\beta ) = \frac{1}{\left[ e^{-\beta J}+2+\sqrt{12-4e^{-\beta J}+e^{-2\beta
J}} \right] }\left( 3+\frac{\left( 12-3e^{-\beta J}\right) }{\sqrt{%
12-4e^{-\beta J}+e^{-2\beta J}}}\right)
\end{equation*}
The case $J > 0$ is not covered by theorem 3.1. From the expression above we
obtain 
\begin{equation*}
l(\beta \rightarrow \infty)=\frac{3+\sqrt{12}}{2+\sqrt{12}} 
\end{equation*}

The only cases where one has analytical solutions are for $q<4$. For other
values of $q$ one has to employ numerical methods in order to gain some
information, as we discuss in the next section.

\section{Numerical analysis}

For those cases which are not analytically treatable we can employ an
algorithm that combines a routine of numerical derivation with eigenvalues
computing. To see how the method work, we first consider the following
matrix, written as a function of $\xi$: 
\begin{equation*}
M(\xi )=\left( 
\begin{array}{ccccc}
e^{-\beta \ J_{0}} & e^{-\beta \ \xi /2} & e^{-\beta \xi } & \cdots & 
e^{-\beta \xi (q-1)/2)} \\ 
e^{-\beta \xi /2} & e^{-\beta \ (J_{1}+\xi )} & e^{-3\beta \xi /2} & \cdots
& e^{-\beta \xi q/2} \\ 
e^{-\beta \xi } & e^{-3\beta \xi /2)} & e^{-\beta \ (J_{2}+2\xi )} & \cdots
& e^{-\beta \xi (q+1)/2} \\ 
\vdots & \vdots & \vdots & \ddots & \vdots \\ 
e^{-\beta \xi (q-1)/2} & e^{-\beta \xi q/2} & e^{-\beta \xi (q+1)/2} & \cdots
& e^{-\beta \ (J_{q-1}+(q-1)\xi )}%
\end{array}%
\right)
\end{equation*}%
Let $\lambda _{\xi }$ be the largest eigenvalue corresponding to matrix $%
M(\xi )$ and $\lambda _{-\xi }$ the one from $M(-\xi )$. If $\xi <<1$ one
has: 
\begin{equation*}
\left. \frac{\partial \lambda }{\partial \xi}\right\vert _{D=0}=\frac{%
\lambda _{\xi }-\lambda _{-\xi }}{2\xi }+\sum\limits_{k=1}^{\infty }\frac{%
\xi ^{(2k+1)}\lambda ^{(2k+1)}(0)} {2(2k+1)!}
\end{equation*}%
since%
\begin{equation}
\begin{array}{ccc}
\lambda _{_{\xi }} & = & \lambda _{0}+\lambda _{0}^{\prime }\xi +\frac{\xi
^{2}}{2}\lambda _{0}^{\prime \prime }+ \frac{\xi^3}{3!}\lambda _{0}^{\prime
\prime\prime } +... \\ 
&  &  \\ 
\lambda _{_{-\xi }} & = & \lambda _{0}-\lambda _{0}^{\prime }\xi +\frac{\xi
^{2}}{2}\lambda _{0}^{\prime \prime }- \frac{\xi^3}{3!}\lambda _{0}^{\prime
\prime\prime } + ...%
\end{array}
\label{eq1}
\end{equation}%
A numerical estimate to order ($\ O(\xi ^{2})\ $) is 
\begin{equation*}
\Delta ^{(1)}\lambda (\xi )=\frac{\lambda _{\xi }-\lambda _{-\xi }}{2\xi }.
\end{equation*}
A more refined numerical estimate can be obtained using not only two but
four points $\lambda _{\xi }$, $\lambda _{-\xi },$ $\lambda _{2\xi }$ and $%
\lambda _{-2\xi }$.

\begin{figure}[th]
\centerline{\psfig{file=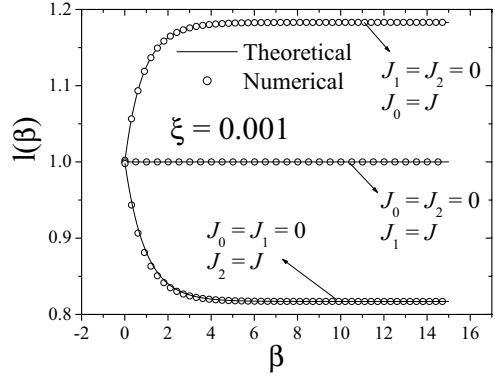,width=8cm}} \vspace*{8pt}
\caption{{}Investment as function of $\protect\beta$ to $q=3$. A comparison
of the numerical result with analitical results.}
\end{figure}

\begin{theorem}
Consider a function 
$\lambda \in C^{\infty }(\mathbb{R})$. A numerical approximation to order $%
O(\xi ^{4})$ of $\lambda ^{\prime }(D)$ is given by
\begin{equation*}
\Delta ^{(2)}\lambda (\xi )=-\frac{1}{3}\Delta ^{(1)}\lambda (2\xi )+\frac{4%
}{3}\Delta ^{(1)}(\xi )
\end{equation*}
\end{theorem}

\begin{proof}
Considering a Taylor expansion
\begin{equation}
\lambda _{2\xi }=\lambda _{0}+2\xi \lambda _{0}^{\prime }+4\xi ^{2}\frac{%
\lambda _{0}^{\prime \prime }}{2!}+8\xi ^{3}\frac{\lambda _{0}^{(3)}}{3!}%
+16\xi ^{4}\frac{\lambda _{0}^{(4)}}{4!}+....  \label{eq3}
\end{equation}%
and
\begin{equation}
\lambda _{2\xi }=\lambda _{0}-2\xi \lambda _{0}^{\prime }+4\xi ^{2}\frac{%
\lambda _{0}^{\prime \prime }}{2!}-8\xi ^{3}\frac{\lambda _{0}^{(3)}}{3!}%
+16\xi ^{4}\frac{\lambda _{0}^{(4)}}{4!}+....  \label{eq4}
\end{equation}

Combining the equations (\ref{eq1}), (\ref{eq3}) and (\ref{eq4}),
we obtain
\begin{equation*}
8\lambda _{\xi }-\lambda _{2\xi }+\lambda _{-2\xi }-8\lambda
_{-\xi }=12\xi \lambda _{0}^{\prime }-\frac{2}{5}\xi ^{5}\lambda
_{0}^{(5)}
\end{equation*}%
So
\begin{equation*}
\lambda _{0}^{\prime }=\left[ \frac{\lambda _{2\xi }-\lambda _{-2\xi }}{%
4\xi }\right] -\frac{4}{3}\left[ \frac{\lambda _{\xi }-\lambda _{-\xi }}{%
2\xi }\right] +O(\xi ^{4})
\end{equation*}%
which gives us
\begin{equation*}
\Delta ^{(2)}\lambda (\xi )=-\frac{1}{3}\Delta ^{(1)}\lambda (2\xi )+\frac{4%
}{3}\Delta ^{(1)}(\xi )\qquad
\end{equation*}
\end{proof}

To assess the applicability and performance of the method, we applied it to
the integrable $q=3$ case with numerical results using 4--point derivative.
In Fig. $3$ we show how the numerical results compare with the analytical
ones. By using a higher number of points we observe a significant difference
on the results (see Fig. $4$) over selected regions, as compared to a lesser
number.

\begin{figure}[th]
\centerline{\psfig{file=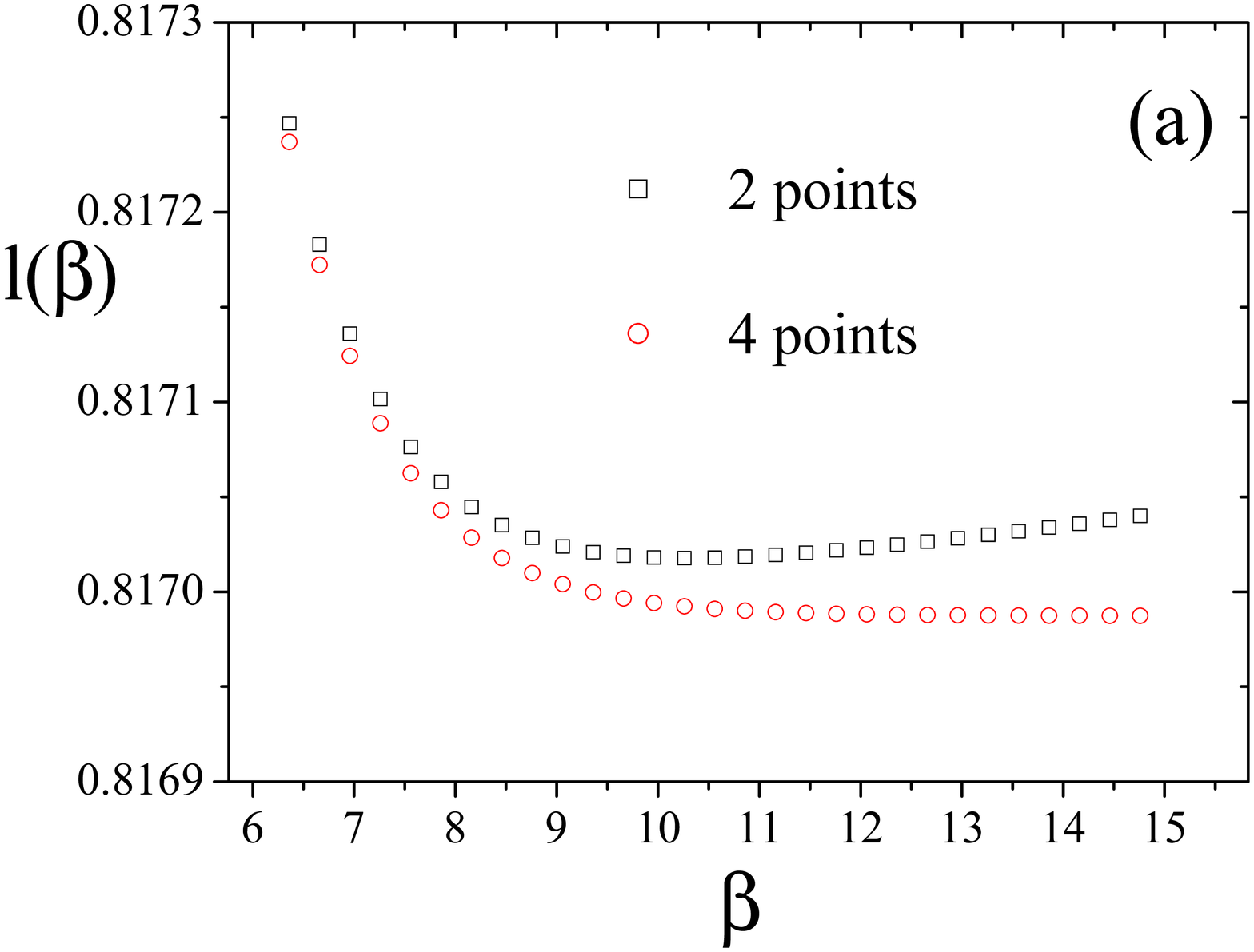,width=8cm}} \centerline{%
\psfig{file=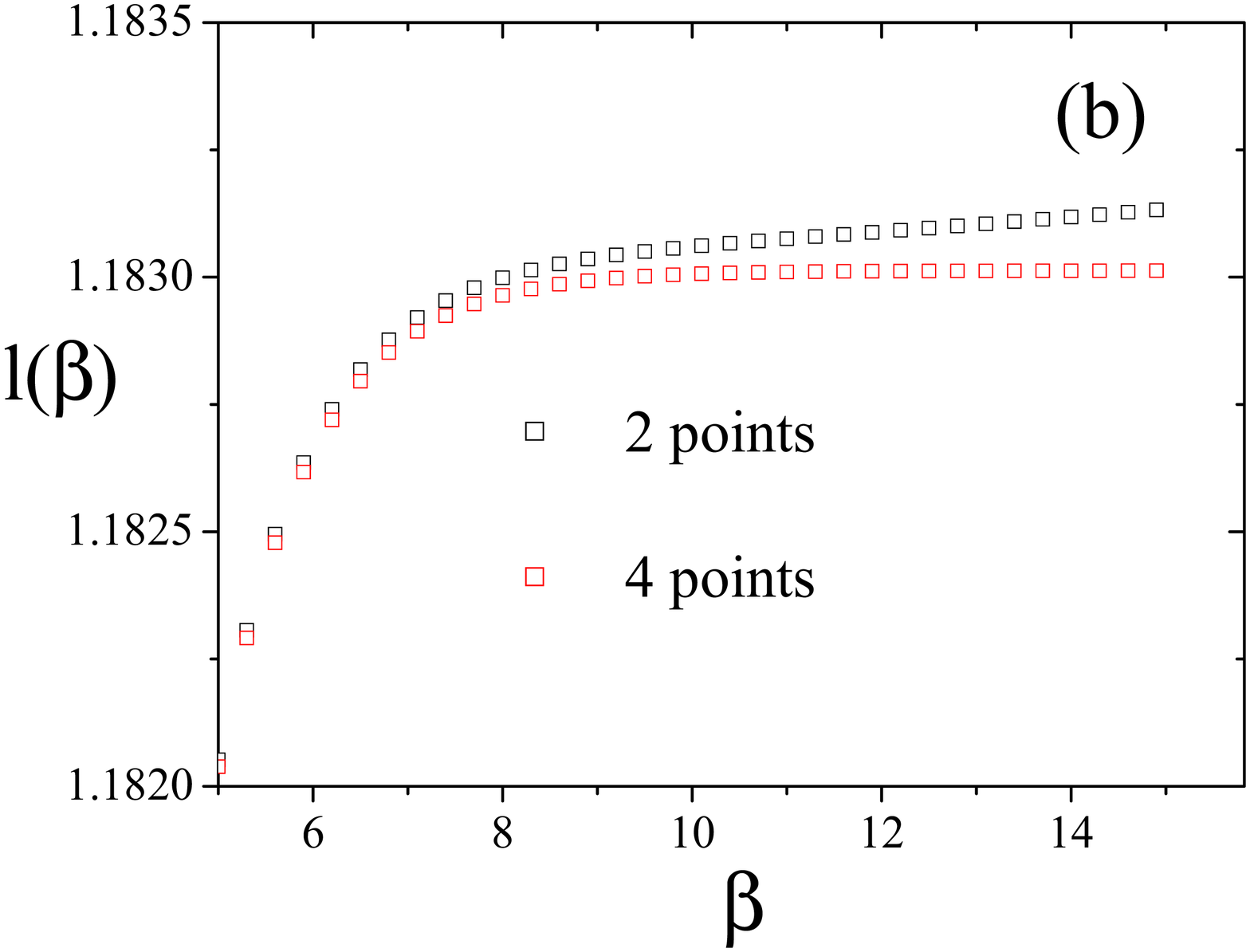,width=8cm}} \centerline{%
\psfig{file=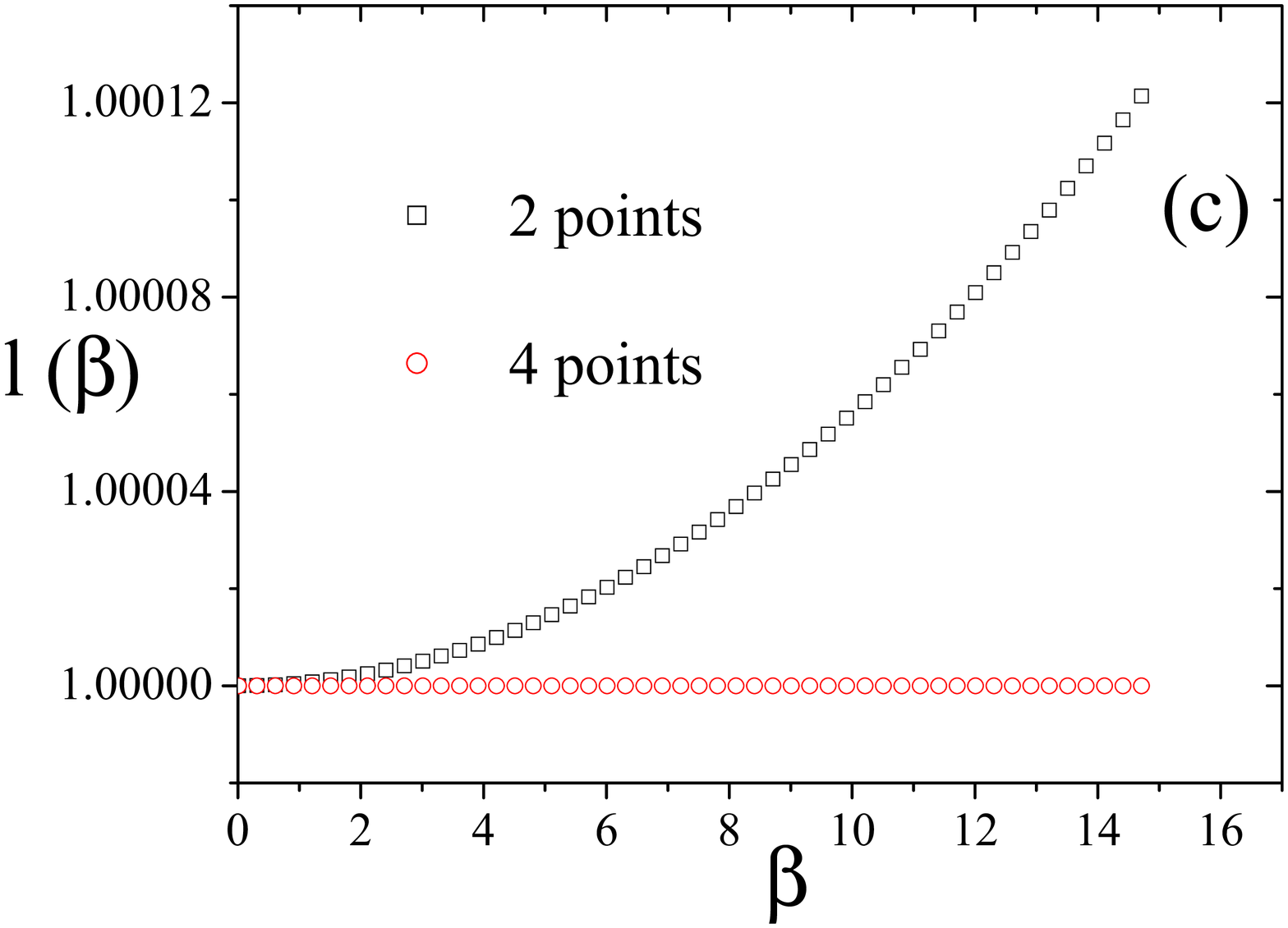,width=8cm}} \vspace*{8pt}
\caption{{}Comparative numerical derivates for $q=3$, using three points and
five points. (a) $J_{0}=J_{1}=0$ and $J_{2}=J$; (b) $J_{1}=J_{2}=0$ and $%
J_{0}=J$; (c) $J_{0}=J_{2}=0$ and $J_{1}=J$. }
\end{figure}

\subsection{Numerical analysis for $q>3$ considering distinct profiles of
agents}

In this section we analyze some numerical results for $q>3$. We consider
three possible profiles:

\begin{enumerate}
\item \textbf{aggressive or risk-prone agents:} in this case the probability
of an agent's investment increase as a function of invested value. We
modelled this through 
\begin{equation*}
J(\sigma _{i})=-(\sigma _{i}+1)\ <0,
\end{equation*}%
where $\sigma _{i}=0,...,q-1$.

\item \textbf{conservative agents:} the probability of investment decreases
as function of invested value. In this situation 
\begin{equation*}
J(\sigma _{i})=-(q-\sigma _{i})\ <0,
\end{equation*}%
where $\sigma _{i}=0,...,q-1$.

\item \textbf{random agents:} the probability of the investment is randomly
chosen for each agent, such that 
\begin{equation*}
J(\sigma _{i})=\left\lfloor \func{rand}[0,1]\cdot q\right\rfloor
\end{equation*}

where$\ \func{rand}[0,1]$ is a random number uniformly generated in the
interval $[0,1]$.
\end{enumerate}

\begin{figure}[th]
\centerline{\psfig{file=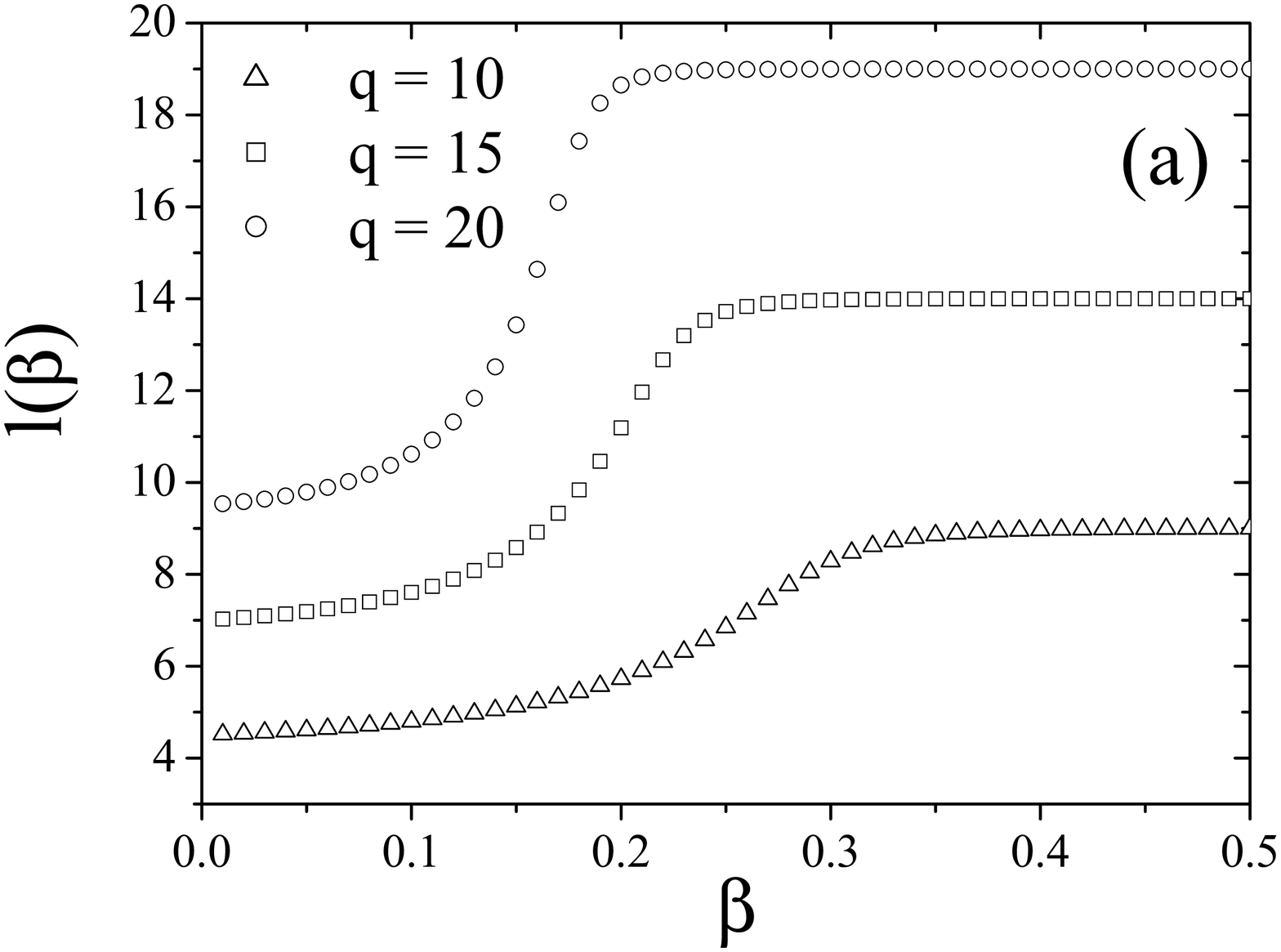,width=8cm}} \centerline{%
\psfig{file=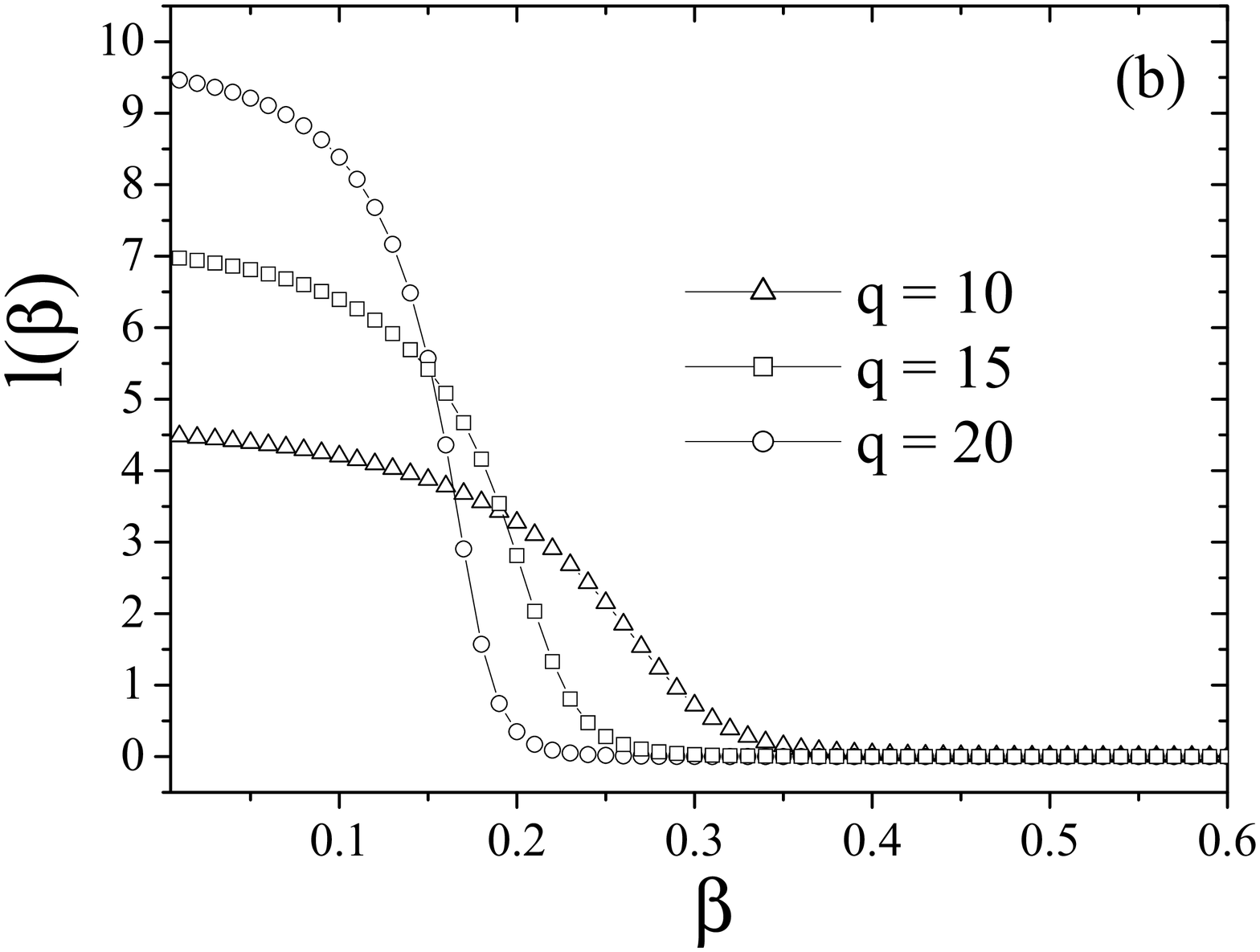,width=8cm}}
\caption{Numerical results for the behavior of investiment as function of $%
\protect\beta $ for $q=10$, $q=15$ and $q=20$ for two diferent profiles: (a)
agressive and (b) conservative. }
\end{figure}

We generated plots with $q=10,15,20$ for three different profiles. In Fig.
5, we show the risk--prone (a) and conservative (b) profiles. From (a) we
conclude that all agents are inclined to invest the maximum quantity for $%
\beta \rightarrow \infty $ since greater quantities are privileged by the
probability distribution. Differently, for (b) as $\beta \rightarrow \infty
\ $, the investment of each agent goes to $0$. We then may conclude that
conservative agents lead to the situation of complete stagnation as $\beta
\rightarrow \infty \ $, independent of the number possibilities in the
investment $q$. On the other hand, risk prone agents lead the market to
invest the maximum at this limit.

An alternative profile seems to be more appropriate: The random profile (3)
was also explored in an experiment using $12$ seeds (12 different random
choices of the string $J(\sigma _{i})$, $i=1...N$) and $q=15$ as depicted in
Fig. $6$ ). This figure represents the average over the 12 seeds. The
behavior of seeds are not similar in the sense that they may yield different
values of investment at $\beta \rightarrow \infty $. 
\begin{figure}[th]
\centerline{\psfig{file=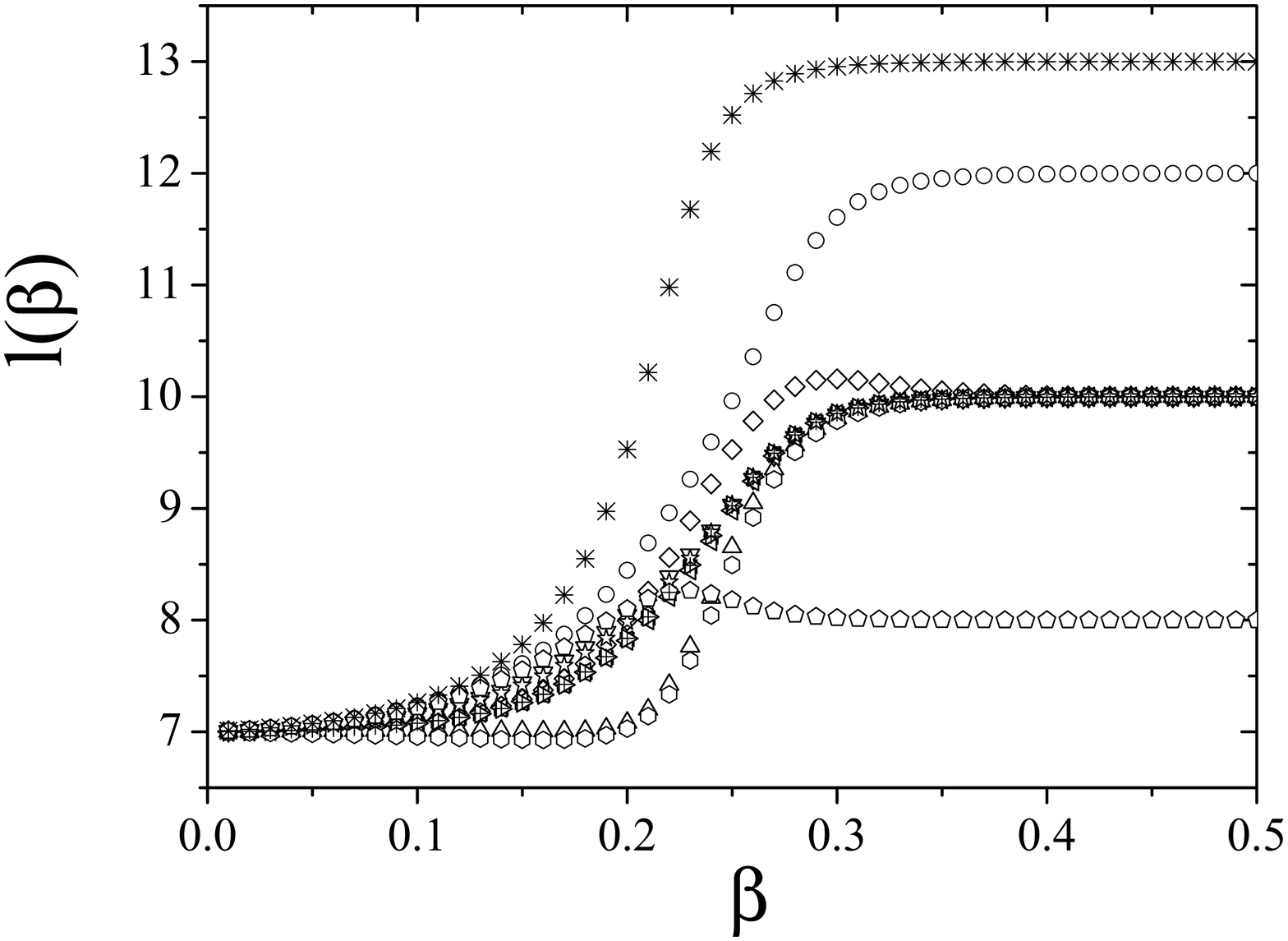,width=8cm}} \centerline{%
\psfig{file=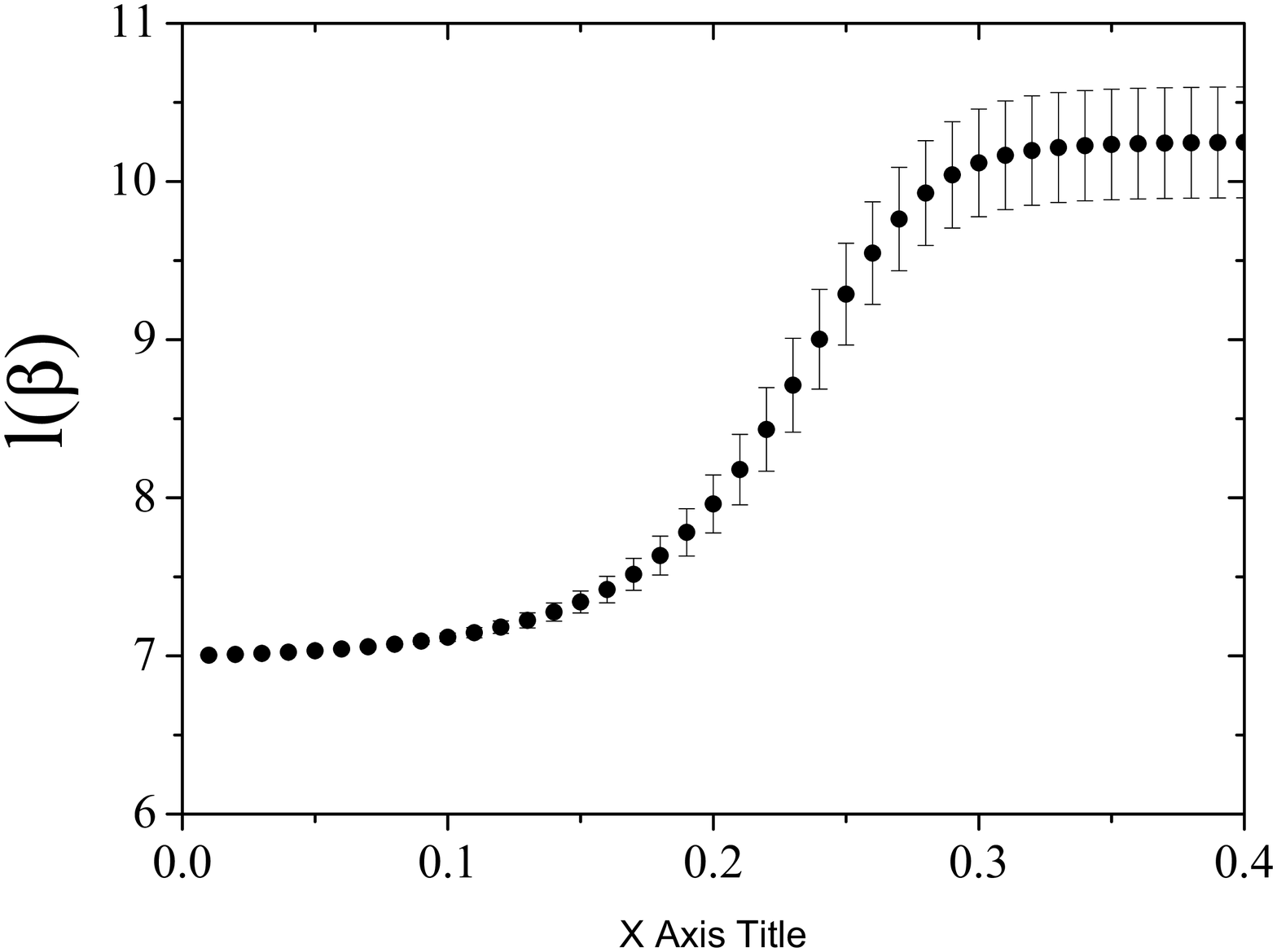,width=8cm}}
\caption{(a) Plot for 12 different seeds for $q=15$; (b) mean value over the
seeds.}
\end{figure}

\section{Summary and conclusions}

We have studied the investment behavior of a group of agents as a function
of a parameter that mimics the mean profit obtained by agents. Our results
illustrated different situations based on possible investors' profiles. In a
model where $q$ represents the number of possible investment amounts, we
obtained analytical results for $q=2$ and $3$. For larger values of $q$ we
performed a series of numerical simulations by combining exact
diagonalization algorithms with numerical derivatives.

Our results indicate that the behavior of each investor is key to
determining the dynamics of the market. As recent results in the context of
agents' simulation show \cite{SAAS2004}, pure mathematical models can
capture some of the intrincacies of real markets when they, as pointed out
in \cite{D1999}, try to include real people's idiosyncrasies (beliefs,
sentiments, etc.) that are known to play a significant role (not to mention
other important influences as seasonable changes in production, political
turmoil, and the like). Even though our model is still mathematical, in the
sense that it is based on a well known model of statistical mechanics and we
identify behavior in terms of known physical quantities, we believe that our
ideas might help indicate a way towards a more realistic market modelling.

\section*{Aknowledgments}

The authors thank A. C. Bazzan for enlightening discussions

\end{document}